# Combined AES/Factor Analysis and RBS Investigation of a Thermally Treated Pt/Ti Metallisation on SiO$_2$

U. Scheithauer, W. Hösler and R. Bruchhaus

Siemens AG, Corporate Research and Development, Otto-Hahn-Ring 6, W - 8000 München 83, FRG

**Abstract:**
Pt/Ti metallisation bilayers are used as bottom electrodes for ferroelectric thin films. During deposition of the ferroelectric films, these electrodes are exposed to elevated temperatures causing modifications of the Pt/Ti bottom electrode. Diffusion and oxidation of the Ti adhesion layer have been studied by the application of factor analysis to AES depth profile data and by RBS.

Factor analysis was employed to extract the chemical information from the measured AES spectra and to derive semiquantitative depth profiles of the identified material compounds. RBS was used to obtain the quantitative depth distribution of the elements. By the combination of both methods, diffusion and oxidation processes were observed and could be precisely describe.

## Introduction:

Pt/Ti bilayer metallisations for use as bottom electrodes for ferroelectric thin films have been investigated after deposition and additional annealing by Auger electron and Rutherford backscattering spectroscopies (AES & RBS).

Especially if the samples are rotated during sputter errosion [1, 2], AES depth profiling is able to probe the samples with a good depth resolution. Conventional peak-to-peak height analysis of the derivative spectra gives an approach to the depth distributions of pure elements. Application of factor analysis (FA) to the data overcomes this limitation [3, 4] and enables an estimate of the in-depth distributions of present compounds.

In contrast to the electron spectroscopy methods, the intensities obtained by RBS are independent on the chemical environment of the elements. Due to this feature and its inherent easy quantifiability, RBS complements depth profiling AES to give a maximum of quantitative compositional information on thin film systems.

In the following, results of both methods are compared and it is demonstrated that the combination of AES/FA and RBS gives a better insight into the diffusion and oxidation processes of the Pt/Ti bilayer which occur during annealing in an $Ar/O_2$ atmosphere.

## Experimental:

The electrode consisting of a 140 nm Pt / 100 nm Ti bilayer was deposited by RF sputtering onto a Si-wafer covered with appoximately 500 nm thermally grown oxide [5]. To study the effect of depositing ferroelectric $Pb(Zr,Ti)O_3$ thin films on this electrode [6], the deposition process was simulated by the heat treatment of the electrode in an $Ar/O_2$ atmosphere. The details of this treatment are given in table 1.

The AES depth profiles were acquired using a Perkin-Elmer PHI-600 scanning Auger instrument equipped with a differentially pumped ion gun. For sputter erosion of the sample, 3 keV $Ar^+$ ions with an impact angle of $71^o$ relative to the surface normal were employed. To ensure a good depth resolution the samples were continuously rotated around the surface normal during sputter erosion [1, 2]. AES data were taken from an area of approximately $5 * 10^4 \mu m^2$ using a primary electron energy of 10 keV and a primary current of 0.7 μA. Sputtering was interrupted during AES data acquisition.

By assuming that each spectrum recorded at a certain sputter depth is a linear superposition of basic spectra, the acquired data could be evaluated by FA [3, 4]. From the recorded data, the multivariant statistical method FA is able to extract both the basic spectra and the corresponding concentrations. FA recognizes the chemical environment of the atoms from altered AES peak positions and peak shapes which themselves represent the density of electronic states and cross sections. So not the in-depth profile of the elements but that of the compounds was determined.

RBS was performed with $^4$He$^{2+}$ projectiles of 2.7 MeV at a backscattering angle of 170°. The RBS data were evaluated by RUMP [7, 8] using the tabulated stopping powers of Ziegler [9]. The simulation procedure for the evaluation of the measured spectra was started from the qualitative layer models proposed by AES/FA. The peak shapes of the RBS signals from Ti and Pt were evaluated. Due to the change of the total stopping power, the shapes of these signals are sensitive to the absolute content and distribution of O within the electrode layer. The nearby resonance in the $^4$He scattering cross section at 2.484 MeV [10] and the uncertainty introduced by substracting the Si background, results in the direct evaluation of the RBS oxygen signal to be less precise. Nevertheless, by use of the cross section given by Leavitt et al. [10], the O signal was also included into the evaluation procedure and gave no hints for discrepancies in the resulting layer models.

## Results and Dicussion:

Fig. 1 shows the results of FA applied to the AES depth profile data. By FA evaluation of the AES depth profile data the basic compounds $SiO_2$, Pt, metallic Ti, slightly oxygen influenced Ti (labeled by Ti′), and Ti in the two different oxidation states $TiO_x$ and $TiO_y$ were found.

In fig. 2 the recorded RBS spectra are compared to the simulated spectra of the best-accordance layer models as depicted in fig. 3. These models are shown before convolution with the instrument resolution function, the energy straggling and the layer thickness spread.

This must be taken into account if the models from the RBS simulations are to be compared with the AES/FA material compound depth distributions. Also, due to the higher depth resolution of AES sputter depth profiling, the FA depth profiles depict additional thin layers. For instance, the $TiO_x$ at the electrode/$SiO_2$ interface of sample A and more fine structures within the layers themselves were detected. Additionally, one has to keep in mind that RBS yields fully quantitative depth distributions of the elements, while the chemically sensitive AES/FA estimates the depth profiles of the material compounds. Considering these differences, the results from both analytical techniques are comparable and complementary.

The most striking differences between the AES/FA depth profiles (Fig. 1) of the three samples can be attributed to the diffusion of Ti into the Pt and towards the surface. This is directly mirrored in the RBS spectra (Fig. 2) which additionally confirm that neither Pt nor Ti was lost during annealing. The broadening of the Ti and Pt signals in the RBS spectra and in the AES/FA depth profiles show that the layer has become thicker during annealing due to the uptake of O and the oxidation of Ti. As confirmed by the measurements, the electrode/$SiO_2$ interface width also increases with annealing time, thereby indicating an increasing lateral non-uniformity of the electrode.

Both methods confirm, AES/FA directly in the depth profile and RBS by differences in the low energy slopes of the signals from Ti and Pt, that even for sample C a Pt-free $TiO_2$ layer remains at the interface to the $SiO_2$.

RBS estimates a Ti/O ratio close to $TiO_2$ for sample C. Attributing this ratio to the material compound labeled $TiO_y$ in the AES/FA depth profiles (Fig. 1) and using the AES peak heights, the material compound $TiO_x$ of sample A can be approximated by TiO.

## Summary:

Combining the analytical capabilities of AES/FA and RBS gives insight into the diffusion and oxidation processes of Pt/Ti electrodes annealed in an $Ar/O_2$ atmosphere. AES sputter depth profiling and FA evaluation of the data yields the semiquantitative depth profiles of the material compounds. These results are complemented by RBS which provides the necessary quantitative compositions and thicknesses.

Annealing the Pt/Ti electrodes in an $Ar/O_2$ atmosphere causes a diffusion of Ti towards the surface and an oxidation of Ti. This process ends with the complete oxidation of Ti to $TiO_2$ and the uptake of O increases the electrode thickness. For the investigated electrode, even after the employed annealing process, a Pt-free $TiO_2$ layer remains at the interface to the $SiO_2$.

| Sample | Treatment |
|---|---|
| A | bottom electrode "as sputtered" |
| B | 20 min ramped in Ar / $O_2$ (1.4 Pa, 1:1) to $T \approx 450$ °C |
| C | 20 min ramped in Ar / $O_2$ (1.4 Pa, 1:1) to $T \approx 450$ °C and 120 min annealed under these conditions |

Table 1: Sample annealing conditions

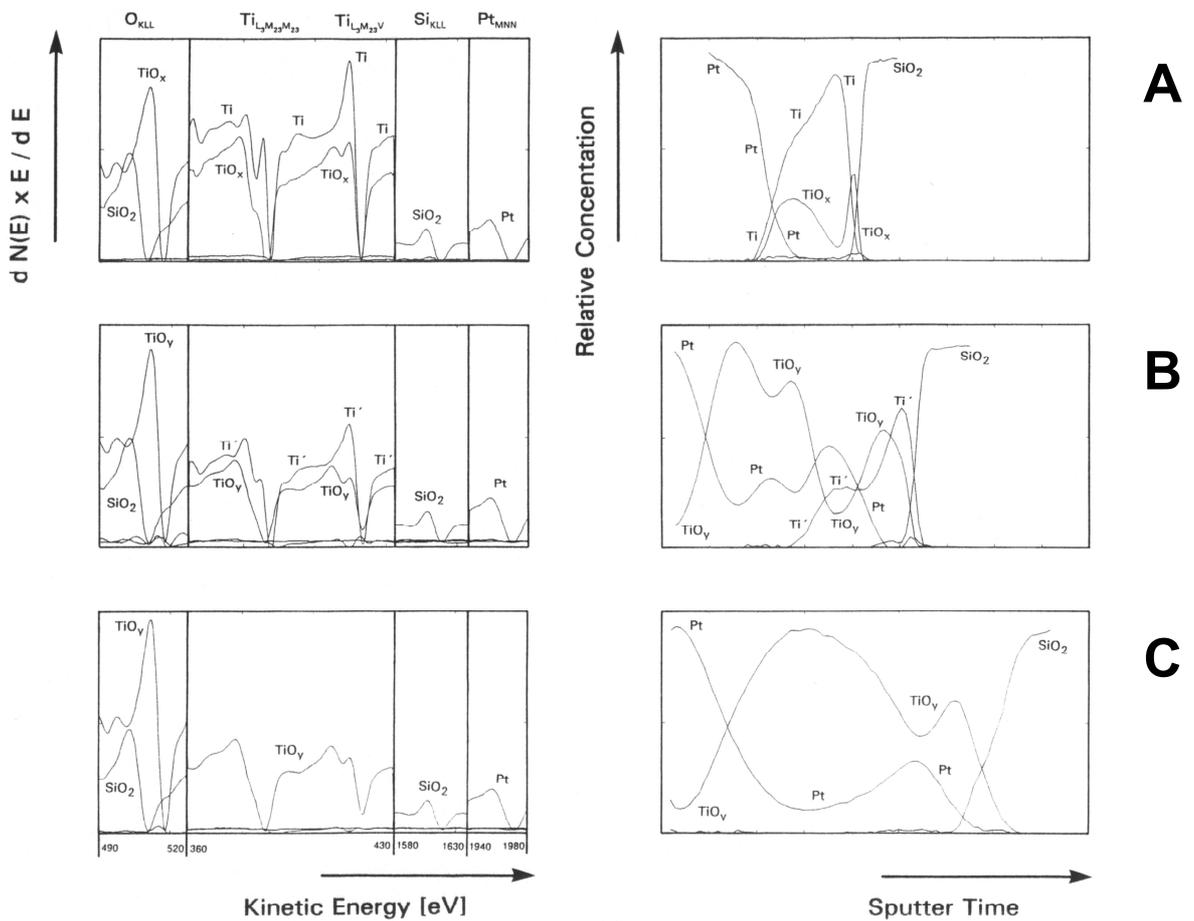

Fig. 1: Results of FA applied to the AES depth profile data. Material compound depth profiles are shown on the right hand side with the corresponding basic spectra on the left. The letters A, B, C refer to the sample preparation as defined in table 1.

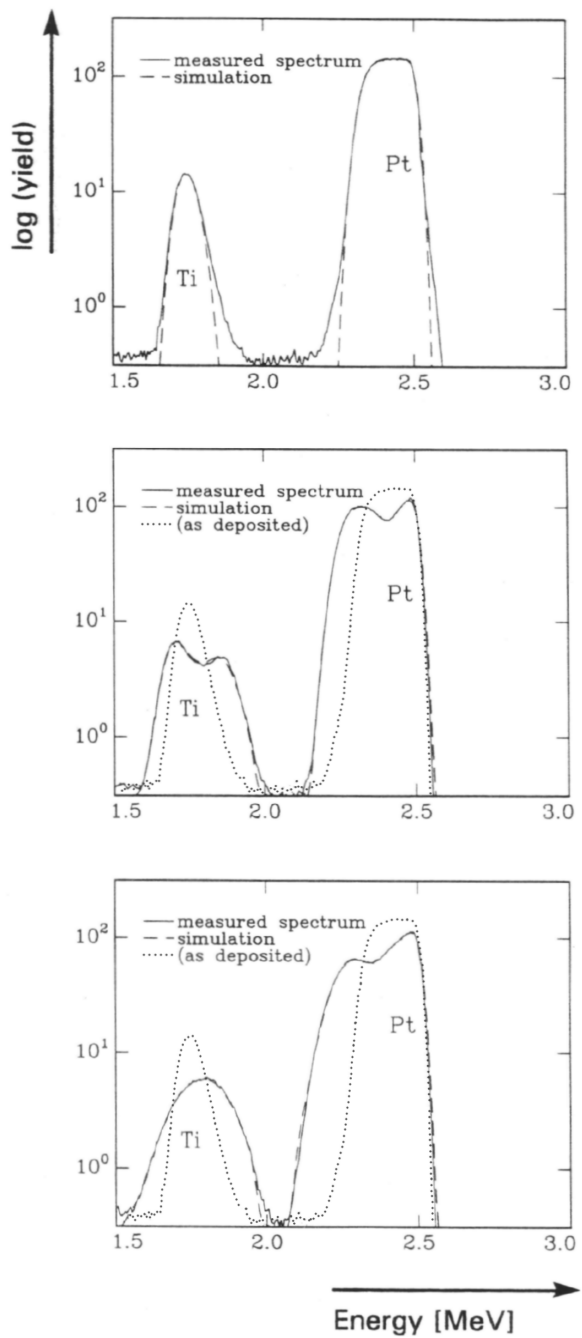
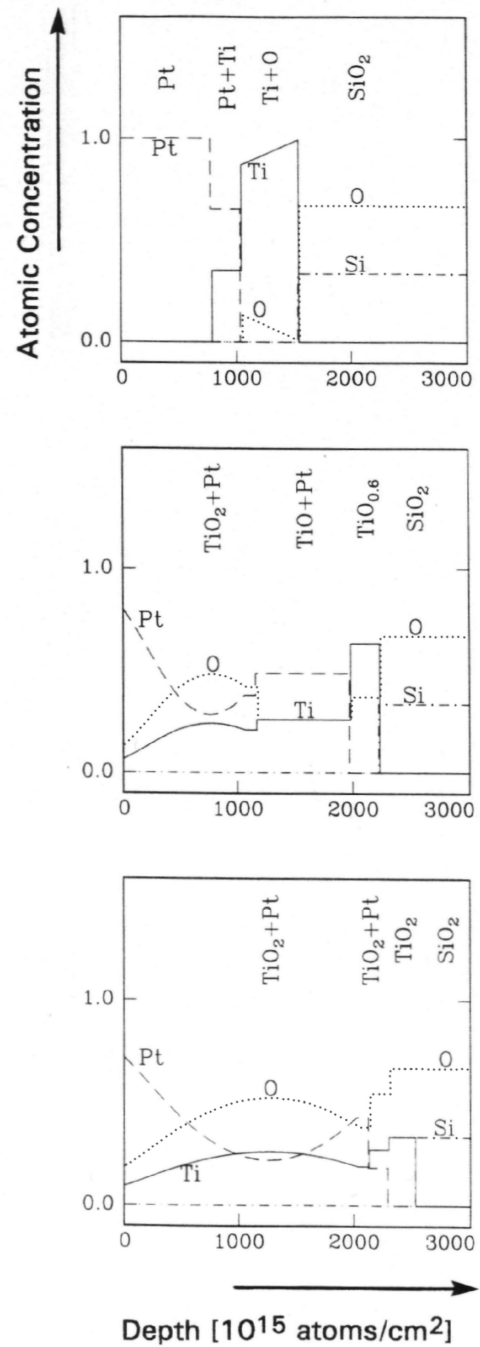

Fig. 2: Measured RBS spectra and the results from the simulation of the models of fig. 3. For comparison, the spectrum from sample A is overlaid onto the spectra from B and C.

Fig. 3: Samples composition models used for the simulations shown in fig. 2.